# Non-contact luminescence lifetime cryothermometry for macromolecular crystallography

V. B. Mykhaylyk[1]*, A. Wagner[1], H. Kraus[2]

[1]Diamond Light Source, Harwell Campus, Didcot, OX11 0DE, UK
[2]University of Oxford, Department of Physics, Denys Wilkinson Building, Keble Road, Oxford, OX1 3RH, UK

**Abstract**

Temperature is a very important parameter when aiming to minimize radiation damage to biological samples during experiments that utilise intense ionising radiation. A novel technique for remote, non-contact, *in situ* monitoring of the protein crystal temperature has been developed for the new I23 beamline at the Diamond Light Source, a facility dedicated to macromolecular crystallography (MX) with long-wavelength X-rays. The temperature is derived from the temperature-dependant decay time constant of luminescence from a minuscule scintillation sensor (<0.05 mm$^3$) located in very close proximity to the sample under test. In this work we present the underlying principle of cryogenic luminescence lifetime thermometry, discuss the features of the detection method, the choice of temperature sensor and demonstrate how the temperature monitoring system was integrated within the viewing system of the end-station used for the visualisation of protein crystals. The thermometry system was characterised using a $Bi_4Ge_3O_{12}$ (BGO) crystal scintillator that exhibits good responsivity of the decay time constant as function of temperature over a wide range (8 – 270 K). The scintillation sensor was calibrated and the uncertainty of the temperature measurements over the primary operation temperature range of the beamline (30 – 150 K) was assessed to be ±1.6 K. It has been shown that the temperature of the sample holder, measured using the luminescence sensor, agrees well with the expected value. The technique was applied to characterise the thermal performance of different sample mounts that have been used in MX experiments at the I23 beamline. The thickness of the mount is shown to have the greatest impact upon the temperature distribution across the sample mount. Altogether these tests and findings demonstrate the usefulness of the thermometry system in highlighting the challenges that remain to be addressed for the in-vacuum MX experiment to become a reliable and indispensable tool for structural biology.

*corresponding author e-mail: vitaliy.mykhaylyk@diamond.ac.uk, phone: +44(0)12358801

# 1. Introduction

Temperature is a crucial parameter that defines the state of a system. Measuring temperature accurately and reliably is thus very important when monitoring chemical, physical and biological processes. As such, its precise and accurate determination is of fundamental importance in science and technology. Specific requirements for temperature monitoring in harsh and/or hardly accessible environments has prompted the development of several non-contact methods for temperature measurements, exploiting a change of optical properties, i.e. emission intensity, refractive index, wavelength shift, luminescence decay time, etc. with temperature (see e.g. review [1] and references therein). These methods are tailored to specific experimental environments and rely on a change of the selected property with temperature but come with their specific shortfalls or drawbacks. Intensity-dependant measurements require stability to maintain calibration, while thermal changes of spectral features are often quite small, resulting in low sensitivity. Furthermore, the range of applications a particular method is suitable for can be rather limited. For example, a most popular and widespread method based on infrared thermometry cannot be used to measure cryogenic temperatures due to the signal (blackbody radiation emitted by an object) dropping significantly with lowering the temperature. Background noise of typical detectors used in these applications therefore renders measurements below ~200 K unfeasible [2].

Of particular interest in this regard is a technique that utilises the temperature variation of the luminescence decay time constant, i.e. luminescence lifetime thermometry. The temperature dependence of the luminescence decay time constant has been known since the beginning of last century but it was not until the late 1970s that advances in electronic instrumentation enabled reliable measurements of the luminescence decay. From then onwards, luminescence lifetime thermometry became the norm in experimental practice and it is also used in a range of applications where more conventional thermometry is difficult to implement, i.e. furnaces of all sort, chemically aggressive environments, turbines, magnetic resonance imaging (MRI), radiofrequency (RF) and microwave processes, high-voltage equipment, etc. [3], [4], [5], [6].

The primary advantage of luminescence lifetime thermometry is that it enables accurate measurements of temperature of a remote object using a small sensor whilst not interfering with the object itself. The sensor can be attached to the measured object and the temperature of the whole assembly remains in equilibrium during the measurement. This feature is essential for cryogenic experiments where accuracy of temperature determination can be significantly affected by heat load through contact wires and thermal resistance of interfaces [7], [8], [9], [10] which are difficult to account for in a reliable manner. Hence, the real temperature at the point of interest can be quite different from the sensor reading [11], [12], [13]. If instead the luminescence response of the sensor is measured in the experiment, the accuracy of the temperature measurement is far less influenced by heat leaks, detail of heat transfer at the interfaces or other effects, giving more reliable estimations of the true sample temperature. This method is even more advantageous when there is a need to measure the temperature of small objects (<1 mm) as the size of the temperature sensor can be made similarly small.

Based on a technique developed earlier for studies of cryogenic scintillators [14] we developed a method for determining cryogenic temperatures in vacuum utilizing non-contact measurements. Specifically, the goal was to develop an instrument for remote in situ monitoring of the temperature of protein samples at the new I23 beamline at Diamond Light Source, dedicated to macromolecular crystallography (MX) experiments, using low-energy X-rays [15]. The beamline operates in vacuum and accurate knowledge of the sample temperature is very important. There is an increased possibility of protein sample heating by intense X-ray radiation in a vacuum environment, similar to what is observed in cryo-electron microscopy studies [16].

In this article we first describe the features of the technique starting with a brief introduction of the measurement principles. We then discuss the characteristics of the components, including choice of sensors, and practical implementation of the system in the suite of beamline instrumentation. The system performance and application examples are then be described and discussed to demonstrate the potential of the developed technique for non-contact, *in situ* measurements of cryogenic temperatures.

## 2. The method

The absorption of high-energy photons in dielectric crystals results in promotion of electrons to the conduction band and formation of holes in the valence band. Such excited states may exist for a certain periods of time but eventually the system returns to its equilibrium state when electrons recombine with the holes. The relaxation of charged particles from the excited to the ground state can occur via two possible competitive processes: either radiative with emission of photons or non-radiative when energy is dissipated as heat i.e. transferred to the crystal lattice as phonons. Emission of photons gives rise to luminescence. In the simplest two-level system modelled as ground and excited state, the population of the excited state $N$ as a function of time $t$ follows from solving:

$$\frac{dN}{dt} = -(w_r + w_{nr})N(t), \tag{1}$$

for which the solution is:

$$N(t) = N_0 \exp(-(w_r + w_{nr})t), \tag{2}$$

where $N_0$ is initial population and $w_r$ / $w_{nr}$ are rates for radiative and non-radiative processes. The emission lifetime $\tau$ is defined as the inverse of the sum of the rates:

$$\tau = \left(w_r + w_{nr}\right)^{-1} \tag{3}$$

The rate associated with the non-radiative processes ($w_{nr}$) exhibits a strong temperature dependence, contributing to a variation of the decay time constant with temperature. This dependence is the key aspect of luminescence lifetime thermometry. The simplest case is the thermal quenching of the luminescence when electrons can be promoted over the energy gap $\Delta E$ to the non-emitting state and then recombine non-radiatively with holes at the expense of photon emission. This results in a decrease of the measured luminescence decay time constant with temperature $T$ as is presented by Mott's formula:

$$\tau^{-1} = \tau_r^{-1} + K \exp(-\Delta E / kT), \tag{4}$$

where $\tau_r = w_r^{-1}$ is the radiative decay constant, $K$ is the non-radiative decay rate and $k$ is the Boltzmann constant. The temperature dependence of the decay time constant given by this equation is modelled in Figure 1 (curve a) using values of $\tau_r$, $K$ and $\Delta E$ close to those observed in typical self-activated scintillation materials [17]. The figure illustrates that for a two-level system the temperature-dependent de-excitation channel dominates at high temperatures. At low temperatures, when $kT \ll \Delta E$, the luminescence decay constant is practically temperature-independent. Effectively, for such a system, the value of the thermal activation energy governing the de-excitation process defines the lower limit of the temperature range where the decay time constant varies enough to be useful for temperature monitoring.

In some materials the features of energy structure of emission centres facilitate a complementary mechanism which results in changes of the decay time constant at low temperatures. Of particular interest is the three-level system composed of a ground state and two excited levels separated by a narrow energy gap $D$. The dynamics of transitions between these levels has been analysed in [18] and it has been shown in measurements that the temperature dependence of the decay time constant can be expressed as following:

$$\tau^{-1} = \frac{k_1 + k_2 \exp(-D/kT)}{1+\exp(-D/kT)} + K\exp(-\Delta E/kT) \qquad (5)$$

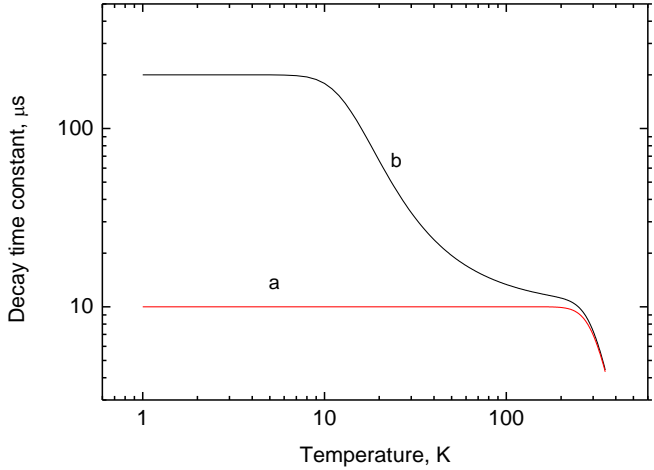

Fig. 1. Luminescence decay time constants as function of temperature, calculated using the two models discussed in the text. The parameters for the two-level model (a, Equation (4)) are: $\tau_r = 10^{-5}$ s, $K = 10^8$ s$^{-1}$ and $\Delta E = 0.2$ eV. The parameters for the three-level model (b, Equation (5)) are: $k_1 = 5\times 10^3$ s$^{-1}$, $k_2 = 2\times 10^5$ s$^{-1}$, $D = 0.005$ eV $K = 10^8$ s$^{-1}$ and $\Delta E = 0.2$ eV.

where $k_1$ and $k_2$ are the rates of radiative decay from the two excited levels. Figure 1 (curve b) shows the temperature dependence of the decay time constant calculated according to equation (5). At very low temperatures the emission is dominated by radiative transition from the lower-lying metastable level and the decay time constant is relatively large. When the temperature increases, the electrons can be promoted to the upper excited level and the overall rate of radiative recombination increases, resulting in a decrease of the measured decay time constant. As is demonstrated by curve b in Fig. 1 this effect causes the luminescence decay time constant to exhibit a significant variation over a wide temperature range. This results in adequate sensitivity for temperature measurements down to temperatures of a few kelvins.

## 3. General approach and system configuration

The equipment for luminescence lifetime thermometry usually consists of the following components: i) a sensor sample, exhibiting a sensible change of the decay time constant over the temperature range of interest, ii) an excitation source, iii) an optical system to deliver excitation and to collect the luminescence signal, iv) a detector, v) a data acquisition (DAQ) system and data analysis package. For the luminescence lifetime thermometry system to be beneficial and practically useful it is vital to attain the best match of all components to ensure efficient excitation, light collection, and signal detection. The emission spectra and decay time constant of the luminescence sensor drive the main requirements on the spectral and timing characteristics of the system components.

The optical components (lenses, optical fibres, light guides) used to collect luminescence light and convey it to the detector are commonly available. The light can be detected by a conventional Si-based diode or a photomultiplier tube (PMT) with broad spectral sensitivity and timing resolution adequate for lifetime measurements. Until recently, the pulsed excitation source (laser or Xe-lamp) and the fast transient recorder were the most costly parts of the equipment that, to some extent, precluded wider use of this technique. Therefore, all previous developments of decay time thermometry have been driven by medicine and technologically important applications, such as power generation or aerospace where cost is a secondary issue. Recent commercialisation of bright LEDs with emission wavelengths reaching the UV-region brought about an affordable and attractive alternative to the laser excitation sources. The cost of fast, general-purpose DAQ systems also dropped. However, the specialised DAQ modules for measurements of fast transient processes remain fairly expensive and often require even more costly customisation of data processing and analysis. That is why development of a specialised luminescence lifetime thermometry system remains a fairly involved task and it is therefore very important to utilise all possibilities for adaptation of existing hardware components and methods for signal detection and analysis.

*3.1. Detection method*

Multiphoton counting (MPC), originally developed for the characterisation of scintillation materials over a wide temperature range [19], is well-suited for determining luminescence decay time constants. MPC is ideal for measurements of decay time constants in the range $10^{-6}$-$10^{-3}$s, which are typically observed in many luminescence materials.

A compact DAQ system has been designed and manufactured, based on a 14-bit ADC, clocked at a sampling frequency of 200 MHz (5 ns sampling interval). A XILINX Spartan-6 FPGA interfaces the ADC to an optical fibre link. A separate adaptor converts the optical fibre media to a standard USB-2 interface. Typical recordings of scintillation events contain a few tens of photons within a short time interval, depending on the decay time constant (see fig.2).

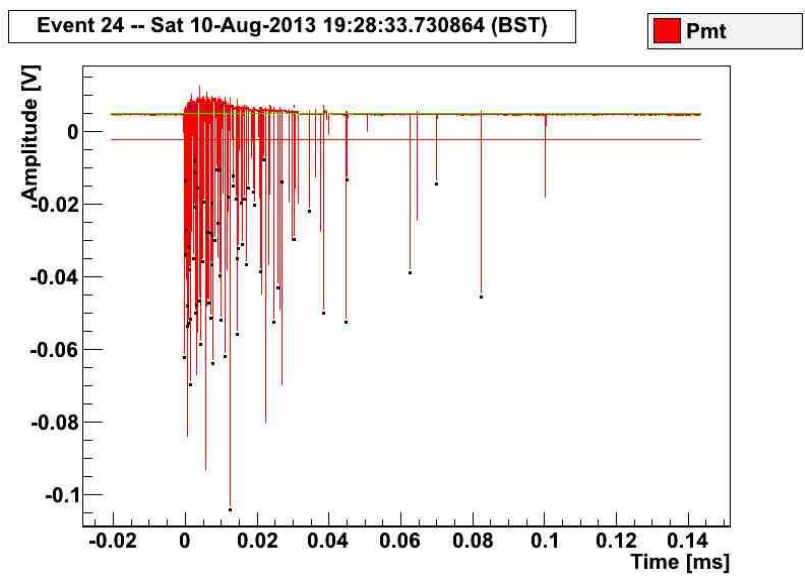

Fig.2. Scintillation event from CaWO$_4$ recorded by DAQ and displayed by data analysis software. The event contains 82 photons.

The analysis software identifies individual photons in the event and derives the arrival time for each photon in the event. The distribution of photon arrival times provides information on the decay characteristics of the emission process. A histogram of photon arrival times that shows a decrease of the emission intensity after excitation (decay curve) can be fitted with the sum of a few exponential functions (Fig.3). In this way, a decay curve of a luminescence process can be recorded and the luminescence decay time constant can be determined. The MPC technique provided an important stepping-stone that underpinned the swift development of the luminescence lifetime thermometry system for the I23 beamline.

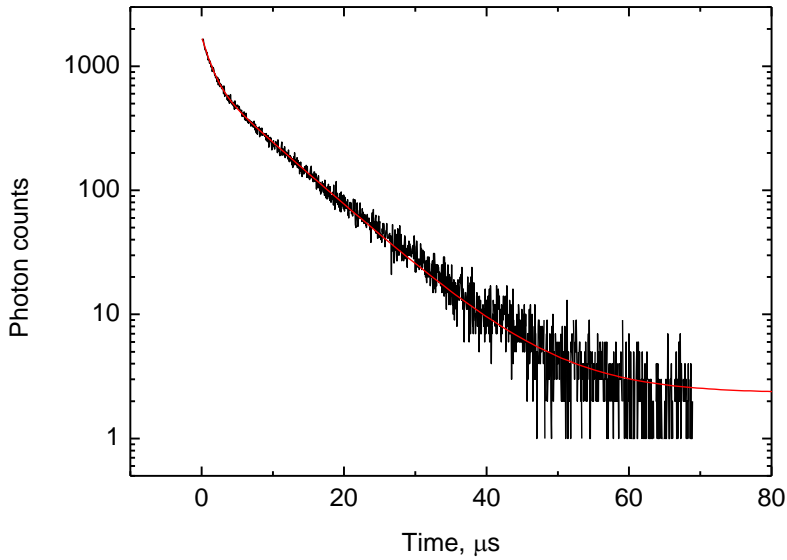

Fig.3 Scintillation decay curve of a CaMoO$_4$ crystal excited by 5.5 MeV alpha-particles emitted by an $^{241}$Am measured at room temperature. The curve is fitted by two exponential functions $y = A_f \exp(-t/\tau_f) + A_s \exp(-t/\tau_s) + y_0$ where $\tau_f = 1.32 \pm 0.01$ and $\tau_s = 8.57 \pm 0.05$ μs.

*3.2. Scintillation crystal as a temperature sensor*

Historically the main research efforts have been directed towards the development of luminescence lifetime thermometry for application at ambient and high temperatures. Various materials doped with rare earth elements, Mn$^{4+}$ and Cr$^{3+}$ [3], [4], [5], [20], [21], [22] as well as diamond particles [23] have been described as luminescence sensors, suitable for measurements in this temperature domain. The application of this technique for the measurements at low temperatures is relatively new. The indication of this is that in a recent review [5], that collated data for 100 phosphors employed for the luminescence lifetime thermometry, only five compounds are mentioned that can be used for measurements below 100 K. Two phosphors, namely Mg$_4$FGeO$_6$:Mn and La$_2$O$_2$S-Eu, are known to meet the requirements for measurements of temperatures down to 10 K [24]. There have been a few feasibility studies of luminescence lifetime thermometry in cryogenic environments. An optical fibre equipped with luminescence sensors (La$_2$O$_2$S-Eu, CaF$_2$-Yb and SrF$_2$-Yb) was successfully used to monitor remotely the temperature of tanks with liquid propellants [25] and of components of cryogenic turbo pumps for rocket engines [26]. However, to the best of our knowledge, no application of scintillation materials has been considered for luminescence lifetime cryothermometry so far.

The choice of material for the sensor is guided by a number of requirements of which the most important is sensitivity in the temperature range of interest. The sensitivity of the luminescence sensor is mostly governed by the steepness of the decay time constant versus temperature curve. Recent characterisation of scintillation materials [27] identified sensors suitable for luminescence lifetime thermometry at cryogenic temperatures. It should be noted that scintillators produce intense luminescence at any high-energy excitation including UV-light, so they are ideal for this application. Moreover, in a few so-called self-activated scintillators, the centre responsible for emission has two excited levels and that causes notable changes of the decay time constant with temperature as presented by curve b in Fig. 1. Some of these compounds like $CaMoO_4$ [28] or $CaWO_4$ [18] have a very steep section making them well suited for a specific, limited temperature range, while others demonstrate good sensitivity over a broader range of temperatures, for instance $Bi_4Ge_3O_{12}$ (well-known under the abbreviated name BGO) [29]. The luminescence of these scintillators is efficiently stimulated by UV radiation at 250 nm which can be provided by modern UV LEDs.

The next criterion for selecting a luminescence sensor is the value of the decay time constant at very low temperatures, i.e. when emission occurs only from the metastable level. It is preferable to have this parameter at $\leq 10^{-3}$ s as the signal-to-noise ratio decreases when the decay time constants of the luminescence process being used become longer. This is due to the emitted photons being spread over a larger time interval and the fraction of noise events increasing. Also it limits the maximum frequency of a pulsed excitation source that, in the end, defines the measurement time. Regarding this criterion, BGO, exhibiting a long decay time constant (ca. 140 µs) is preferred over $CaWO_4$ and $CaMoO_4$. Thus, we made our final choice in favour of BGO based on the suitability for the specific application i.e. monitoring temperature of a protein sample holder in vacuum environment.

*3.3. System components and operation*

A specific characteristic of the I23 beamline is that in order to eliminate adverse absorption and scattering of low-energy X-ray photons all components of the experimental end station are placed in a vacuum vessel with a pressure $<10^{-7}$ mbar[1]. A sample of a protein crystal is attached to the sample holder which is then mounted in the receptacle of a cryogenic goniometer. The goniometer provides centring, orientation and rotation of the sample during diffraction experiments. It also ensures sample cooling through a system of flexible thermally conductive links connected to the pulse tube cooler. Although the temperature of the goniometer receptacle is monitored by a platinum resistor thermometer, the presence of a breakable joint, a few interfaces and, above all, a poorly thermally conductive thin dielectric sample mount leads to substantial temperature differences across the sample holder assembly shown in fig. 4. Consequently, the real temperature of the protein sample is difficult to establish with any meaningful certainty. It can be significantly higher than that of the goniometer receptacle.

---

[1] For a more detailed description of MX experiments in a vacuum at the I23 beamline, the reader is referred to the recent paper [35]. Here we discuss only the issues related to the subject of present study.

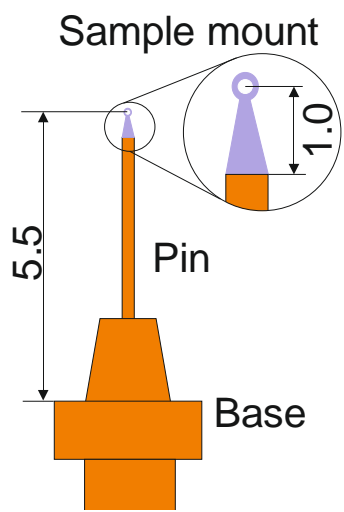

Fig. 4. Schematic of the sample holder assembly for mounting of protein crystals (units are in mm). The holder contains a magnetic base, a pin (both made of copper) and a sample mount which is made of thin dielectric film (10 – 20 μm) with low X-ray absorption/scattering.

To address this critical issue, a non-contact luminescence thermometry system was designed and added to the instrumentation of the I23 end-station. The strategy was to integrate some bulky hardware components of the thermometry (excitation source, optics and photodetector) in the pre-existing modules or systems. In this regard the high-performance optical system for visualisation of the protein crystals is an indispensable part of equipment used in MX experiments. Since for temperature measurements the luminescence sensor should be in the position of the sample or very close, the viewing system itself can be used to collect and transport the luminescence light to the photodetector. The use of the existing viewing system offered a very simple and elegant way for integration of the temperature monitoring system into the end-station equipment. It required only minimal modification to the optics, i.e. merely addition of a motorised mirror that allowed light to be directed towards a photodetector. A schematic of the thermometry system integrated into the on-axis viewing system is shown in Fig. 5 (top). To switch from sample visualisation to thermometry mode, mirror M3 can be inserted into the light path and which then redirects the photons towards the photodetector. The view from the top inside the vacuum vessel in Fig. 5 (bottom) shows the goniometer with the mounted sample and front lens of the viewing system with the UV LED attached.

An excitation source UV LED (model UVTOP240, Qphotonics) with ball lens is attached to the housing of the front lens of the viewing system and light emitted from it is directed towards the sample. The LED is driven by a waveform generator and produces pulses of excitation light with $\lambda = 245$ nm ($\Delta\lambda=12$ nm) at a repetition frequency of 200 Hz. The excitation frequency is chosen to ensure that the time interval between two consecutive pulses exceeds five times the value of the longest decay time constant observed in the luminescence sensor. The same signal from the pulse generation provides a trigger for the MPC.

The MPC method of detection described in section 3.1 requires a photodetector capable of counting individual photons in a luminescence event. A conventional photo multiplier tube (PMT) with high gain and low jitter is currently the detector of choice for this application. A PMT with bialkali photocatode (Model 9107B from ET Enterprises) was integrated into the viewing system of the experimental end-station. The signal from the PMT is transmitted through vacuum-compatible cabling and a vacuum feedthrough to the outside of the vacuum vessel and fed into the fast DAQ system. The digitised signal is then transferred via a fibre-optic cable to a receiver with USB converter and fed to the computer. The power to the PMT is supplied by a stabilised source outside of the vacuum vessel.

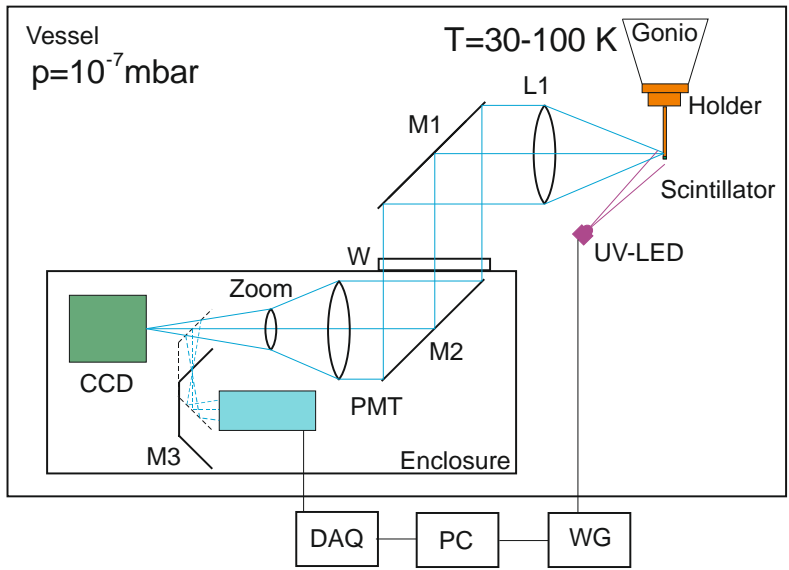

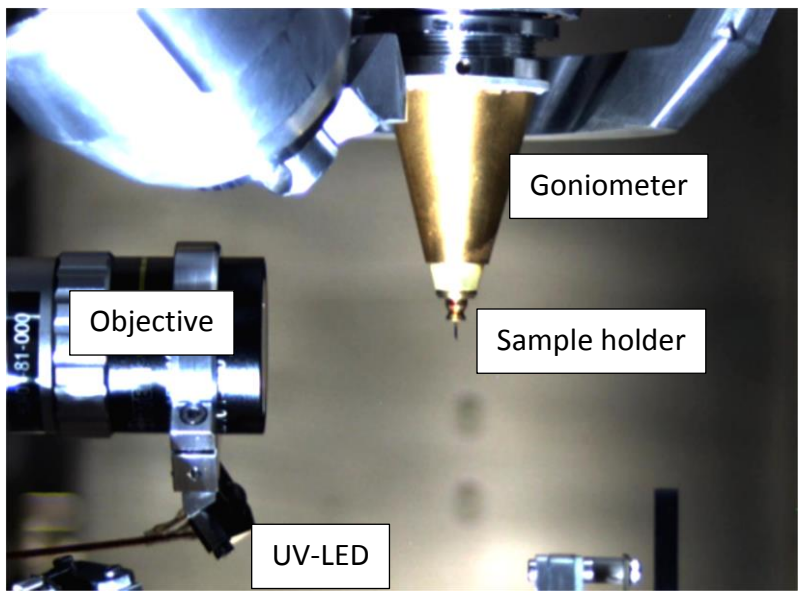

Fig. 5 (top) Schematics of the luminescence lifetime thermometry system: M1-M3 mirrors, L1 – objective lens UV LED – ultraviolet light emitting diode, W – optical window, PMT – photomultiplier tube, CCD – camera, DAQ – data acquisition, PC – personal computer, WG - waveform generator. (bottom) View of the in-vacuum sample environment from the top showing LED, objective and scintillator sensor mounted in the goniometer receptacle.

For the measurements a splinter of a scintillation crystal (sensor) is affixed to the place of the sample holder assembly where temperature needs to be monitored. The sample holder is then loaded into the vacuum vessel, placed into the receptacle of the cryogenic goniometer and the sensor is centred in the field of view of the optical viewing system using rotation and translation axes of the goniometer. Then the pulsed LED is switched on to start the measurement. In principle one luminescence pulse contains information about the decay time constant. However, to improve accuracy a few thousands of the luminescence cycles excited in the scintillator are recorded and analysed using algorithms of the MPC method. The duration of measurements for one temperature point depends on the length of recorded interval and temperature. Usually it takes from 10 to 30 s to acquire statistically significant datasets that allow the luminescence decay time constant to be determined with an error <1%. The temperature is then derived from the value of the decay time constant using a calibration curve. Real-time temperature measurements are also possible [23] though have not been tried yet as this requires more complex data processing.

### 4. Sensor calibration

For the luminescence sensor to provide a reliable temperature reading it must be calibrated. This requires measurements of the luminescence decay time constant as a function of temperature over the temperature range of interest using a well-characterised reference thermometer. The BGO scintillator used in this study was characterised in an optical He flow cryostat using a calibrated Cernox temperature sensor from Lakeshore. Since the accuracy of the calibration may be affected by factors other than temperature [30] it is important to ensure that the calibration is done under the same conditions as during real measurements. This requirement is rather difficult to satisfy as the setup used for the calibration measurements differs from the thermometry system described in the previous section. Nonetheless, efforts were made to replicate at least the most important parts of the hardware, namely the same excitation source (pulsed UV LED), photodetector (bialkali PMT) and detection technique were used in the calibration measurements. The sensor was mounted on a copper base of the cryostat holder. A plate-like splinter of BGO crystal with linear dimensions under 0.5 mm was attached to the copper base using Apiezon N grease to ensure the best possible thermal contact between the temperature sensor and the measured crystal. The temperature during the measurements was monitored and stabilised by the PID loop of a Lakeshore 331 temperature controller.

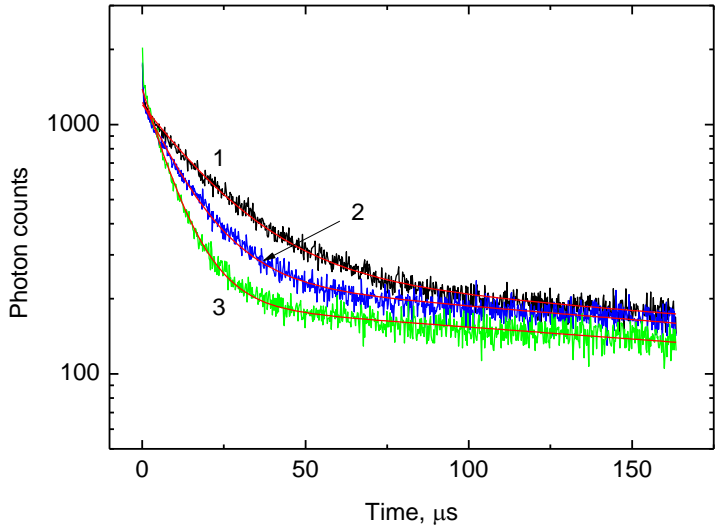

Fig. 6. Luminescence decay curves of the BGO scintillator at UV excitation (λ=245 nm) measured at different temperatures. The red lines show the fitting to the experimental data using two exponential decay functions $y = A_f \exp(-t/\tau_f) + A_s \exp(-t/\tau_s) + y_0$ for different temperatures: 1- T=31 K; $\tau_f$ =20.7±0.19 μs, 2- T=50 K; $\tau_f$ =13.1±0.12 μs, T=121 K; $\tau_f$ =6.8±0.07 μs.

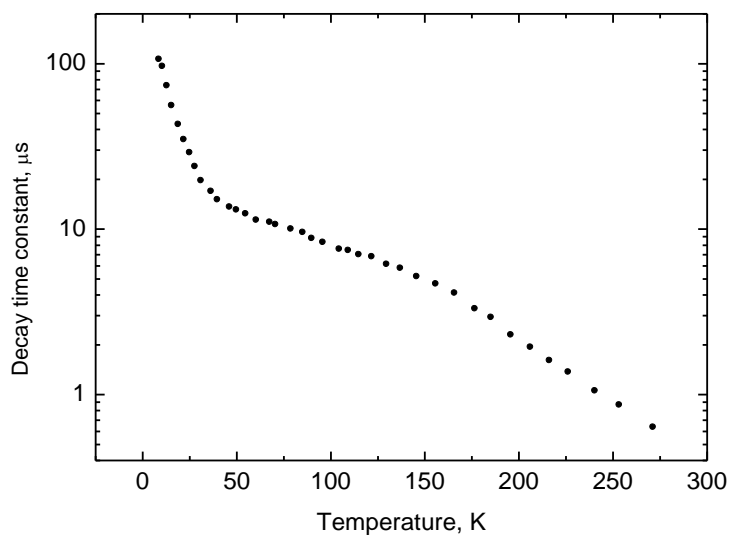

Fig.7 Luminescence decay time constant as function of temperature for the BGO scintillator at UV excitation (λ=245 nm).

Examples of decay curves measured at different temperatures are shown in Fig 6. The decay curve contains fast and slow emission components therefore the fitting to the experimental data was done using the sum of two exponential functions. The fast component due to the intrinsic emission of BGO crystal exhibits clear temperature dependence as is shown in Fig. 7. The long component has extrinsic origin. It does not change with temperature and makes up about 20% of the total intensity of the luminescence signal. This emission is attributed to the spurious luminescence excited by intense UV radiation in the glass of the LED lens or in organic materials (grease, glue, etc.). Either of these may exhibit visible luminescence upon UV excitation [31], [32]. Figure 8 shows such spurious emission (blue) that can be seen around the luminescence sensor emitting cyan light. By carrying out deconvolution analysis of the decay curves a useful luminescence signal with a fast decay time constant can be separated from the spurious emission with the long decay constant (ca. $10^{-3}$ s). Thus, the impact of this spurious emission upon the performance of the non-contact thermometry system is negligible.

The calibration curve in Fig. 7 covers the range from 8 to 270 K but currently, for practical reasons, the temperature range for luminescence lifetime thermometry at the I23 beamline is reduced. The low-temperature limit is determined by the lowest temperature that the goniometer receptacle can reach (30 K). At the other end of the temperature range the upper limit for present applications is set at 150 K. This is dictated by the necessity to keep a sample below the temperature of phase transition from vitreous to crystalline ice, typically occurring above 130 K [33]. When heated above this temperature trapped radicals produced as a result of X-ray absorption during the diffraction experiment become mobile and damage the protein crystals causing the quality of diffraction to deteriorate dramatically.

Since the aim of luminescence lifetime thermometry is to deliver a reading of the temperature from the measured value of the decay time constant the data were plotted as a function $T = f(\tau)$ and were then fitted by a fifth-order polynomial. The main contribution to the error comes from the uncertainty in the fit of the fast decay time constant (±1%). The average sensitivity of a BGO scintillation sensor in the 30 – 150 K range is 0.07 μs/K. Consequently, for a representative value of the measured decay time constant of 10 μs the uncertainty of the fit is ±0.1 μs which translates to a ±1.4 K error in temperature. Combining this with the error of the calibration (±0.8 K) gives the uncertainty of the temperature measurements equal to ±1.6 K. This error is smaller at lower temperatures (<30 K) due to better sensitivity, while in the high temperature range (150 – 250 K) it gradually increases due to the reduction of the signal and an enhanced uncertainty in the determination of the decay constant. Nonetheless, such level of accuracy is similar to what has been reported so far for many other luminescence lifetime thermometry systems [3], [21], [23]. It is also adequate for the application of this thermometry system, intended for non-contact *in situ* monitoring of the temperature of frozen protein samples.

## 5. System testing and application examples of non-contact temperature measurements

To validate the calibration of the BGO sensor we performed several measurements using the beamline thermometry system. All tests were conducted in the absence of X-ray radiation. The contribution of impinging UV radiation to the heating of the sample holder assembly was assessed. In continuous mode, the typical output power of the UV LED is 70 μW, distributed over a spot with ca. 8 mm diameter at the location of the sample. Thus, assuming that the copper pin with diameter 0.5 mm is illuminated and 100% of the UV radiation is absorbed, the delivered power does not exceed a few μW. This is negligible in comparison with the total heat load on the sample holder assembly, the main cause of which is from black body radiation, estimated to be 2 – 5 mW at T=40 K.

Initially we measured the temperature in places where it is known or can be derived with good accuracy. In stationary conditions all parts of the sample holder assembly which are made of highly thermally conductive copper (see Fig.4) remain at the same temperature. Previously it has been shown that the specially developed magnetic mount provides reliable thermal contact and a predictable temperature gradient across the interface between the sample holder and goniometer receptacle [10], [34]. Hence, the temperature of the sample holder assembly can be estimated from the reading of a PT-100 thermometry sensor integrated into the goniometer receptacle, e.g. at 40 K the temperature of sample holder should be ca. 3 K higher.

A chip of BGO sensor was attached by means of Apiezon N grease to the end of the sample holder pin. The sample holder with the mounted sensor was than inserted into the vacuum vessel and placed in the goniometer receptacle cooled to T=38.6 K following the transfer procedure established for protein samples [35]. It usually takes about 10 minutes for the temperature of the goniometer receptacle to stabilise. When the scintillation crystal was illuminated by UV light from the LED, the bright blue-green emission typical for BGO was clearly visible using the beamline viewing system (see Fig. 8). The luminescence decay curves of the BGO crystal were recorded using the thermometry system, analysed and the temperature of the sensor was derived as equal to 41.1±0.9 K. The measured temperature agrees with the expected value that is calculated as the sum of the temperature of the goniometer receptacle and temperature rise at the goniometer-sample holder interface (38.6 K +3 K=41.6 K).

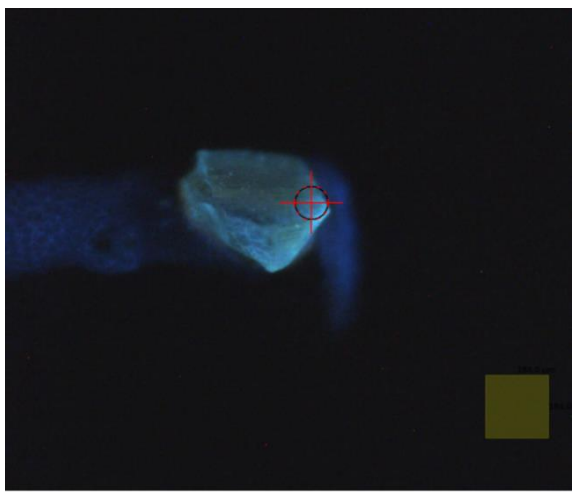

Fig. 8. Photograph of the luminescence excited in the BGO sensor under UV illumination. The chip of the crystal is attached to the copper pin of the sample holder. The square in the corner of the picture has dimensions 184 μm×184 μm.

This test showed that the developed luminescence lifetime thermometry system can provide sensible and fairly accurate measurements of temperature. The main advantages of the system can be recognized when it is used for non-contact determination of temperature across the dielectric sample mount. This part of the sample holder assembly has very poor thermal conductivity due to the nature of the materials used that are dielectrics. Therefore, in the very first series of experiments we conducted measurements of temperature using sample mounts made of different materials. For this test to be representative it was essential to reproduce the environment in which the experiments with the protein crystals are conducted. For this reason, we used a cryoprotectant - 20% water solution of glycerol - a medium which protects protein crystals during freezing and also provides an interface between the frozen crystals and the sample mount. A grain of BGO crystal having the dimensions ca. 100 – 200 μm was immersed in the solution and then placed at the very tip of the mount i.e. the place where protein crystal usually stays during X-ray diffraction experiment (see Fig. 9). Prepared in this manner, samples were flash frozen in liquid nitrogen and transferred to the goniometer receptacle that was kept at a temperature of 40±1 K.

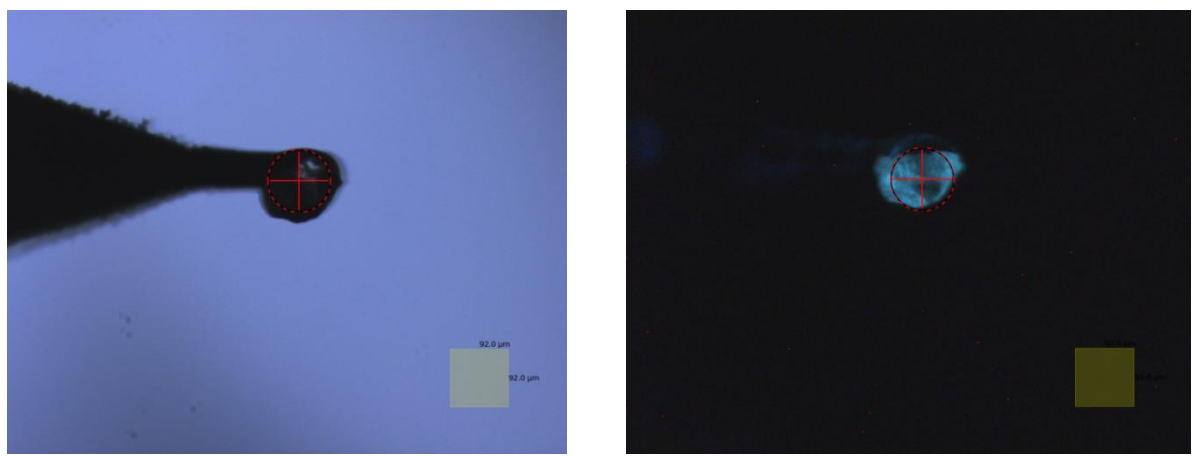

Fig. 9 a) Photograph of the BGO sensor attached to the tip of the sample mount made of 20 μm thick glassy carbon. b) The same as (a) under UV illumination. The squares in the corner of the pictures have dimensions 92 μm×92 μm.

We investigated four types of sample mounts which are currently used to support protein crystals during X-ray diffraction experiments. The results show that there is a significant temperature rise at the tip of the sample mount (see Table 1). The measurements revealed that two types of mount (Sigradur and Litholoop) exhibit the best thermal performance: the temperature at the tip of the mount stayed below 70 K. The temperature of the sensor mounted on Kapton film increased to 93 K while in the worst case (MicroMount) the temperature is 110 K. Interestingly, the first two mounts' types were previously empirically identified to yield the highest fraction of successful X-ray diffraction experiments.

Table 1. Characteristics of investigated sample mounts and results of temperature measurements using the BGO sensor.

| Mount material | Product (Manufacturer) | Thickness, μm | Thermal conductivity, W/mK | | Sensor temperature, K |
|---|---|---|---|---|---|
| | | | 40-100 K | 300 K | |
| Glassy carbon | Sigradur (HTW) | 20 | 0.3-0.7 [36] | 4.6 | 60 |
| Polyimide | Litholoop (Molecular Dimension) | 25 | 0.2-0.6 [37] 0.1-0.2 [7] | | 68 |
| Polyimide | Kapton (Dupont) | 12 | 0.2-0.6 [37] 0.1-0.2 [7] | 0.2-0.4 | 93 |
| Polyimide | MicroMount (Mitegen) | 10 (loop) 25 (stem) | 0.2-0.6 [37] 0.1-0.2 [7] | | 110 |

It should be mentioned that all the tests were done as a demonstration of the technique and that the findings are merely indicative. There are many parameters and factors that influence the heat conductance and subsequently the temperature. These include the properties of the material used (thermal conductivity), the specific geometry (distance, cross-section, profile) as well as the amount and distribution of interposer (cryoprotectant), etc. Nonetheless, even these fragmentary results allowed us to identify important trends. Table 1 includes two known quantities: thickness of the mounts and published data on thermal conductivity of glassy carbon and polyimide. Comparing the data obtained for the sample mounts made of the same material (polyimide) it is fairly clear that the thickness of the film is the main cause of the temperature variation at the sample mount tip. Furthermore, the thickness of the mount in its narrow region (loop area) has a major impact on the heat transport. Concerning the glassy carbon mount we believe that its better performance is due to the fact that thermal conductivity of the material used in this test is higher in comparison with polyamide. This is consistent with the results of recent measurements of the thermal conductivity of amorphous carbon [36].

The temperature rise observed in this experiment has always been expected but the magnitude of the effect has never been quantified. In the next test we endeavoured to measure how temperature changes across the sample mount. For this measurement a sample mount made of Kapton (Dupont) was wetted in the cryoprotectant and the grains of BGO crystal were dispersed across the mount allowing a few grains with the size 50 – 100 μm to stick to the surface. The goniometer translations allowed different parts of the sample mount to be placed in the field of view while the zoom module defined a small region of interest permitting detection of luminescence from the selected grain. The variation of temperature along the sample mount measured in this way is shown in Fig. 9.

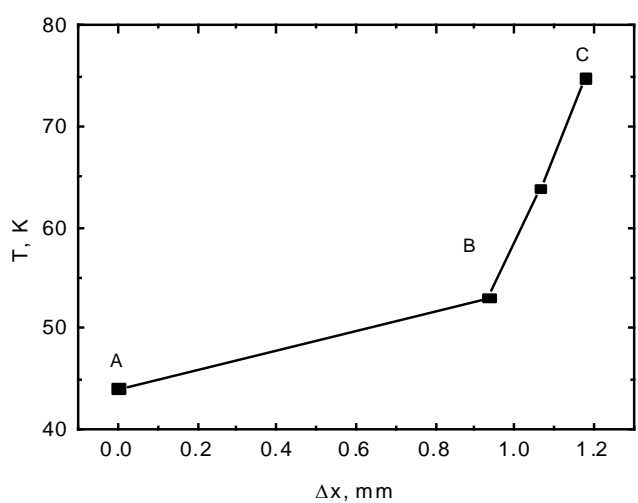

Fig. 10. Variation of temperature across the sample mount made of 12 μm thick Kapton (Du Pont) film. A - copper pin; B - joining of stem with loop, C - tip of the mount. Goniometer temperature T=42 K.

The temperature of the copper pin (point A) is found to be close to the temperature of the goniometer receptacle. It gradually increases when moving to the point where the stem joins with the opening in a loop (point B). At this point the temperature starts to rise rapidly because of the much reduced cross-section of the material in this part of the sample mount. The temperature at the tip of the sample mount was found to be 76 K – less than in the previous test. We explain this difference by the fact that the sample mount was covered by cryoprotectant which facilitated the thermal conductance of the whole assembly. Overall, this experiment demonstrates the potential of the developed non-contact thermometry system for investigations of the temperature distribution with sub-millimetre spatial resolution.

**Conclusions**

Measurement of the temperature of cryogenically cooled samples in a vacuum environment is a challenging task which requires specialist technical solutions. A technique, enabling non-contact *in situ* monitoring of temperature has been developed for the new I23 beamline at Diamond Light Source dedicated to MX experiments using long-wavelength X-rays. The temperature is derived from the luminescence decay constant of a BGO scintillation sensor exhibiting a pronounced variation of the decay time constant over a wide temperature range. One of the main virtues of the non-contact thermometry system is the elimination of all connections between the sensor and the readout system, making it fully compatible with the necessity of swift replacement and manipulation (transfer, mounting, rotation) of the samples. The BGO scintillation sensor was calibrated over the 8 – 270 K temperature range and the accuracy of the measurements was evaluated. In the temperature range of beamline operation (30 – 150 K) the error of temperature determination is found to be ±1.6 K or less.

The measurements of temperature using different configurations and sample mounts were carried out as validation tests for the beamline thermometry system. The temperature of the sample holder measured using the luminescence sensor was found to agree with the expected value. The technique was applied to quantify the thermal performance of different sample mounts that have been used for the MX experiment at the I23 beamline. It has been shown that the magnitude of the temperature rise across the sample holder assembly varies in a wide range, i.e. from 60 to 110 K, while the temperature of the goniometer was 40 K. The obtained results not only explained previous empirical findings but also demonstrated how this technique can aid studies of the complex relationships between various parameters and factors which influence the heat conductance, and subsequently temperature, of the samples. The knowledge gained in such studies is very important for further improvement of sample holder assembly and optimisation of every stage of the sample transfer process that previously has been done by exercising a laborious and time-consuming trial-and-error approach.

**Acknowledgement**

The development of the non-contact luminescence lifetime cryothermometry system for the I23 beamline at Diamond Light Source was supported by the Science and Technology Facilities Council through grant ST/K002929.